\documentstyle[12pt]{article}                    
\newfont{\ffont}{msym10}                        
\newcommand{\beq}{\begin{equation}}             
\newcommand{\eeq}{\end{equation}}               
\newcommand{\bqry}{\begin{eqnarray}}            
\newcommand{\eqry}{\end{eqnarray}}              
\newcommand{\bqryn}{\begin{eqnarray*}}          
\newcommand{\eqryn}{\end{eqnarray*}}            
\newcommand{\preprint}[1]{\begin{table}[t]      
            \begin{flushright}                  
            \begin{large}{#1}\end{large}        
            \end{flushright}                    
            \end{table}}                        
\newcommand{\PD}[2]                             
    {\frac{\partial^{#2}}{\partial #1^{#2}}}    
\begin{document}
\preprint{LA-UR-96-1809 \\ IASSNS-96/47}
\title{Hadronic Resonance Spectrum May Help \\ in Resolution of Meson Nonet
Enigmas}
\author{\\ L. Burakovsky\thanks{Bitnet: BURAKOV@QCD.LANL.GOV} \
\\  \\  Theoretical Division, T-8 \\  Los Alamos National Laboratory \\ Los
Alamos NM 87545, USA \\  \\  and  \\  \\
L.P. Horwitz\thanks{Bitnet: HORWITZ@SNS.IAS.EDU. On sabbatical leave from
School of Physics and Astronomy, Tel Aviv University, Ramat Aviv, Israel.
Also at Department of Physics, Bar-Ilan University, Ramat-Gan, Israel  } \
\\  \\ School of Natural Sciences \\ Institute for Advanced Study \\ Princeton
NJ 08540, USA \\}
\date{ }
\maketitle
\begin{abstract}
The identification of problematic meson states as the members of the 
quark model $q\bar{q}$ nonets by using a hadronic resonance spectrum is 
discussed. The results favor the currently adopted $q\bar{q}$ assignments
for the axial-vector, tensor, and 1 $^3F_4$ $J^{PC}=4^{++}$ meson nonets, and 
suggest a new $q\bar{q}$ assignment for the scalar meson nonet which favors 
the interpretation of the $f_0(980)$ and $f_0(1710)$ mesons as non-$q\bar{q}$
objects. We also suggest that the 2 $^3S_1$ $\frac{1}{2}(1^{-})$ state should 
be identified with the $K^\ast (1680)$ rather than $K^\ast (1410)$ meson.
\end{abstract}

\bigskip
\section*{  }
The existence of a gluon self-coupling in QCD suggests that, in addition to 
the conventional $q\bar{q}$ states, there may be non-$q\bar{q}$ mesons: bound 
states including gluons (gluonia and glueballs, and $q\bar{q}g$ hybrids) and 
multiquark states \cite{1}. Since the theoretical guidance on the properties 
of unusual states is often contradictory, models that agree in the $q\bar{q}$
sector differ in their predictions about new states. Among the naively 
expected signatures for gluonium are \hfil\break
i) no place in $q\bar{q}$ nonet, \hfil\break
ii) flavor-singlet coupling, \hfil\break
iii) enhanced production in gluon-rich channels such as $J/\Psi (1S)$ decay, 
\hfil\break iv) reduced $\gamma \gamma $ coupling, \hfil\break v) exotic 
quantum numbers not allowed for $q\bar{q}$ (in some cases). \hfil\break
Points iii) and iv) can be summarized by the Chanowitz $S$ parameter \cite{Cha}
\beq
S=\frac{\Gamma (J/\Psi (1S)\rightarrow \gamma X)}{{\rm PS} (J/\Psi (1S)
\rightarrow \gamma X)}\times \frac{{\rm PS} (\gamma \gamma \rightarrow X)}{
\Gamma (\gamma \gamma \rightarrow X)},
\eeq
where PS stands for phase space. $S$ is expected to be larger for gluonium 
than for $q\bar{q}$ states. It should be pointed out, however, that mixing
effects and other dynamical effects such as form-factors obscure these
simple signatures. If the mixing is large, only counting the number of 
observed states remains a clear signal for non-exotic non-$q\bar{q}$ states.
Exotic quantum number states $(0^{--},0^{+-},1^{-+},2^{+-},\ldots )$ would be
the best signatures for non-$q\bar{q}$ states. It should be also emphasized 
that no state has unambiguously been identified as gluonium, or as a 
multiquark state, or as a hybrid. In this letter we shall discuss meson states 
whose interpretation as the members of the conventional quark model $q\bar{q}$
nonets encounters difficulties, resulting in the corresponding enigmas 
\cite{enigmas}. We shall be mainly concerned with the scalar, axial-vector, 
and tensor meson nonets which have the following $q\bar{q}$ quark model 
assignments, according to the most recent Review on
Particle Properties \cite{data}:\hfil\break
1) $\; 1\; ^3P_0$ scalar meson nonet, $\;\;J^{PC}=0^{++},\;\;$
$a_0(980),\;\;f_0(975),\;\;f_0(1400),\;\;K_0^\ast (1430)$\hfil\break   
2) $\;1\; ^3P_1$ axial-vector meson nonet, $J^{PC}=1^{++},\;$
$a_1(1260),\;f_1(1285),\;f_1(1510),\;K_{1A}$ \footnote{The $K_{1A}$ is a nearly
$45^o$ mixed state of the $K_1(1270)$ and $K_1(1400)$ \cite{data2}.}\hfil\break
3) $\;1\; ^3P_2$ tensor meson nonet, $\;J^{PC}=2^{++},\;$
$a_2(1320),\;f_2(1270),\;f_2^{'}(1525),\;K_2^\ast (1430),$ and start with the 
discussion of the isoscalar states.
 \\   \\   
1. Scalar meson nonet.\hfil\break
There are four established isoscalars with $J^{PC}=0^{++},$ the $f_0(975),\;
f_0(1400),\;f_0(1590)$ and $f_0(1710),$ and two which need experimental
confirmation, the $f_0(1240)$ and $f_0(1525).$ 
In the quark model, one expects two 1 $^3P_0$
states and one 2 $^3P_0$ $(u\bar{u}+d\bar{d})$-like state below 1.8 GeV. 
Therefore, at least three of the six cannot find a place in the quark model. 
The $f_0(1400)$ and $f_0(975)$ are currently included as two 1 $^3P_0$ states
in the scalar meson nonet. There exists, however, an interpretation of the
$f_0(975)$ as a $K\bar{K}$ molecule \cite{Wei} since it [and the $a_0(980)]$
lies just below the $K\bar{K}$ threshold which is 992 MeV \cite{Flatte}. 
If the $f_0(975)$ is not the 1 $^3P_0$ $s\bar{s}$ state, the latter should be 
found near 1500 MeV with decay widths as expected from flavor symmetry. The 
weak signal as 1515 MeV claimed by the LASS group \cite{LASS} does not have 
the expected large width \cite{Torn90}. In this case, the $f_0(1525)$ could be
a candidate for the 1 $^3P_0$ $s\bar{s}$ state \cite{Mont}. This $f_0(1525)$ 
has been identified as $K\bar{K}$ $S$-wave intensity peaking at the mass of 
the $f^{'}(1525)$ and having a comparable width \cite{Aston,Baub}. The $f_0(
1240)$ is seen in phase shift analysis of the $K_s^0K_s^0$ system \cite{Etkin}.
At present, experimental data are not sufficient to draw firm conclusion on the
nature of this state. The $f_0(1590)$ has been seen in $\pi ^{-}p$ reactions at
38 GeV/c \cite{Binon,Alde87}. It has a peculiar decay pattern for $$\pi ^0\pi 
^0:K\bar{K}:\eta \eta :\eta \eta ^{'}:
4\pi ^0=\;<0.3:\;<0.6:1:2.7:0.8,$$ which could favor a gluonium interpretation 
\cite{Ger}. Another possibility is that it is a large deuteron-like $(\omega
\omega -\rho \rho )/\sqrt{2}$ bound state (``deuson'') \cite{Torn91}. With 
respect to the $f_0(1400),$ a large gluonium mixing is not excluded
because the $\eta \eta /\pi \pi $ branching ratio is only half of the 
flavor-symmetry prediction \cite{Alde86}. The $f_0(1710)$ has been mainly seen
in the ``gluon-rich'' $J/\Psi (1S)$ radiative decay, where it is copiously
produced, and in central production by the WA76 experiment \cite{Armst} at 300
GeV/c $pp$ interactions. It has not been seen in hadronic production $K^{-}p
\rightarrow K\bar{K}\Lambda $ \cite{Aston} nor in $\gamma \gamma $ fusion. Its
$S$ parameter favors a large gluonium component. An attempt to justify the
interpretation of the $f_0(1710)$ as a glueball has been done recently by 
Sexton {\it et al.} \cite{Sex}.
 \\   \\
2. Axial-vector meson nonet.\hfil\break 
The $q\bar{q}$ model predicts a nonet that includes two isoscalar 1 $^3P_1$
states with masses below $\sim $ 1.6 GeV. Three ``good'' $1^{++}$ objects are
known, the $f_1(1285),\;f_1(1420)$ and $f_1(1510),$ one more than expected.
Thus, one of the three is a non-$q\bar{q}$ meson, and the $f_1(1420)$ is the 
best non-$q\bar{q}$ candidate \cite{Cald}. Most likely, it is a multiquark 
state in the form of a $K\bar{K}\pi $ bound state (``molecule'') \cite{Long},
or a $K\bar{K}^\ast $ deuteron-like state (``deuson'') \cite{Torn91}.
 \\   \\ 
3. Tensor meson nonet.\hfil\break
The two 1 $^3P_2$ $q\bar{q}$ states are likely the well-known $f_2(1270)$ and
$f_2^{'}(1525)$ currently adopted by the Particle Data Group \cite{data}, 
although the observation by Breakstone \cite{Break} of the $f_2(1270)$ 
production by gluon fusion could indicate that it has a glueball component. At
least five more\footnote{We do not consider the $f_2(1430),\;f_2(1640),\;
f_2(2150)$ and $f_2(2175)$ states which are included in the recent Meson 
Summary Table \cite{data1} but need experimental confirmation.} $J^{PC}=2^{++}$
states have to be considered: the $f_2(1520),\;f_2(1810),
\;f_2(2010),\;f_2(2300)$ and $f_2(2340).$ Of these, the $f_2(1810)$ is likely 
to be the 2 $^3P_2$, and the three $f_2$'s above 2 GeV could possibly be the 2
$^3P_2$ $s\bar{s}$, 1 $^3F_2$ $s\bar{s},$ and 3 $^3P_2$ $s\bar{s},$ but a 
gluonium interpretation of one of the three is not excluded. The remaining 
$f_2(1520)$ has been rediscovered in 1989 by the ASTERIX collaboration 
\cite{Ast} as a $2^{++}$
resonance in $p\bar{p}$ $P$-wave annihilation at 1565 MeV in the $\pi ^{+}\pi 
^{-}\pi ^0$ final state. Its mass is better determined in the $3\pi ^0$ mode 
by the Crystal Barrel collaboration \cite{CB} to be 1515 MeV, in agreement 
with that seen previously \cite{BG}. It has no place in a 
$q\bar{q}$ scheme mainly because all nearby $q\bar{q}$ states are already 
occupied. Dover \cite{Dover} has suggested that it is a ``quasinuclear'' 
$N\bar{N}$ bound state, and T\"{o}rnqvist \cite{Torn91} that it is a 
deuteron-like $(\omega \omega +\rho \rho )/\sqrt{2}$ ``deuson'' state. 
 \\  \\
Let us now briefly dwell upon another problematic member of the scalar meson 
nonet, the isovector-scalar $a_0(980).$ Its mass, $(982\pm 2)$ MeV, is low, 
compared to its isovector partners, like the $a_1(1260),$ $a_2(1320)$ and $b_
1(1235).$ Its apparent width (as measured in its $\eta \pi $ decay mode), $(54
\pm 10)$ MeV, is small, compared to ist partners (which have 100 MeV and more).
Moreover, neither the relative coupling of the $a_0(980)$ to $\eta \pi $ and 
$K\bar{K},$ nor its width to $\gamma \gamma ,$ are known well enough to draw 
firm conclusions on its nature $(q\bar{q}$, $2q2\bar{q}$ state, $K\bar{K}$ 
molecule, etc.) Another known $1^{-}(0^{++})$ state is the $a_0(1320).$ This 
state, identified as intensity peaking at the mass of the $a_2(1320)$ and 
having a comparable width \cite{Bout}, needs experimental confirmation.

A new candidate, the $a_0(1450),$ has been recently reported by the Crystal 
Barrel Collaboration at LEAR \cite{CB2}. This $a_0(1450)$ is observed in the 
annihilations $p\bar{p}\rightarrow \eta \pi ^0\pi ^0$ at rest in liquid 
hydrogen. One finds that in this particular reaction, the dominant initial 
state is unique, i.e., $^1S_0,$ and in the final state one observes dominantly
processes of the type $0^{-+}\rightarrow 0^{-+}+0^{++},$ like $p\bar{p}
\rightarrow \eta f_0(975),$ $p\bar{p}\rightarrow \pi a_0(980).$ {\it A priori,}
one may suppose that the $a_0(980)$ is the ground state of the $q\bar{q}$
$I=1$ $J^{PC}=0^{++}$ meson, and the $a_0(1450)$ its radial excitation. One
may as well suggest that the $a_0(980)$ is a non-$q\bar{q}$ object and the
$a_0(1450)$ [or $a_0(1320)]$ is the ground state of the $1^{-}(0^{++}).$ In 
this way, an attractive choice for the $q\bar{q}$ scalar meson nonet could be 
the $a_0(1450),$ $K_0^\ast (1430),$ $f_0(1400)$ and $f_0(1525)$ or $f_0(1590),$
as suggested recently by Montanet \cite{Mont}. This choice would leave out the
$a_0(980)$ and $f_0(975)$ which could be then interpreted in terms of 
four-quark or $K\bar{K}$ molecule states, and one may then speculate, with 
some good reasons, that the $f_0(1710)$ is a glueball, or, at least, a state 
rich in glue, in favor of the arguments of Sexton {\it et al.} \cite{Sex}. 
 \\  \\
In this letter we suggest non-traditional approach to the problem of the 
identification of the true $q\bar{q}$ states, viz., the use of a hadronic 
resonance spectrum. It is well known that the correct thermodynamic description
of hot hadronic matter requires consideration of higher mass excited states, 
the resonances, whose contribution becomes essential at temperatures $\sim 
O(100)$ MeV \cite{Shu,Leut}. The method for taking into account these 
resonances was suggested by Belenky and Landau \cite{BL} as considering 
unstable particles on an equal footing with the stable ones in the 
thermodynamic quantities; e.g., the formulas for the pressure and energy 
density in a resonance gas read
\beq
p=\sum _ip_i=\sum _ig_i\frac{m_i^2T^2}{2\pi ^2}K_2\left(\frac{m_i}{T}\right),
\eeq
\beq
\rho =\sum _i\rho _i,\;\;\;\rho _i=T\frac{dp_i}{dT}-p_i,
\eeq
where $g_i$ are the corresponding degeneracies, $$g_i=\left[
\begin{array}{ll}
(2J_i+1)(2I_i+1) & {\rm for\;non-strange\;mesons} \\
4(2J_i+1) & {\rm for\;strange}\;(K)\;{\rm mesons} \\
2(2J_i+1)(2I_i+1)\times 7/8 & {\rm for\;baryons}
\end{array} \right. $$
These expressions may be rewritten with the help of a {\it resonance spectrum,}
\beq
p=\int _{m_1}^{m_2}dm\;\tau (m)p(m),\;\;\;p(m)\equiv \frac{m^2T^2}{2\pi ^2}
K_2\left(\frac{m}{T}\right),
\eeq
\beq
\rho =\int _{m_1}^{m_2}dm\;\tau (m)\rho (m),\;\;\;\rho (m)\equiv 
T\frac{dp(m)}{dT}-p(m),
\eeq
normalized as 
\beq
\int _{m_1}^{m_2}dm\;\tau (m)=\sum _ig_i,
\eeq
where $m_1$ and $m_2$ are the masses of the lightest and heaviest species, 
respectively, entering the formulas (2),(3). 

In both the statistical bootstrap model \cite{Hag,Fra} and the dual resonance 
model \cite{FV}, a resonance spectrum takes on the form
\beq
\tau (m)\sim m^a\;e^{m/T_0},
\eeq
where $a$ and $T_0$ are constants. The treatment of hadronic resonance gas by
means of the spectrum (7) leads to a singularity in the thermodynamic 
functions at $T=T_0$ \cite{Hag,Fra} and, in particular, to an infinite number
of the effective degrees of freedom in the hadronic phase, thus hindering a
transition to the quark-gluon phase. Moreover, as shown by Fowler and Weiner
\cite{FW}, an exponential mass spectrum of the form (7) is incompatible with
the existence of the quark-gluon phase: in order that a phase transition from 
the hadron phase to the quark-gluon phase be possible, the hadronic spectrum
cannot grow with $m$ faster than a power. 

In our previous work \cite{spectrum} we considered a model for a transition 
from a phase of strongly interacting hadron constituents, described by a 
manifestly covariant relativistic statistical mechanics which turned out to be
a reliable framework in the description of realistic physical systems 
\cite{mancov}, to the hadron phase described by a resonance spectrum, Eqs. 
(4),(5). An example of such a transition may be a relativistic high temperature
Bose-Einstein condensation studied by the authors in ref. \cite{cond}, which 
corresponds, in the way suggested by Haber and Weldon \cite{HW}, to 
spontaneous flavor symmetry breakdown, $SU(3)_F\rightarrow SU(2)_I\times 
U(1)_Y,$ upon which hadronic multiplets are formed, with the masses obeying 
the Gell-Mann--Okubo formulas \cite{GMO}
\beq
m=a+bY+c\left[ \frac{Y^2}{4}-I(I+1)\right];
\eeq
here $I$ and $Y$ are the isospin and hypercharge, respectively, and $a,b,c$
are independent of $I$ and $Y$ but, in general, depend on $(p,q),$ where
$(p,q)$ is any irreducible representation of $SU(3).$ Then the only 
assumption on the overall degeneracy being conserved during the transition 
leads to the unique form of a resonance spectrum in the hadron phase:
\beq
\rho (m)=Cm,\;\;\;C={\rm const}.
\eeq
Zhirov and Shuryak \cite{ZS} have found the same result on a phenomenological 
ground. As shown in ref. \cite{ZS}, the spectrum (9), being used in the  
formulas (4),(5) (with the upper limit of integration being infinity), leads to
the equation of state $p,\rho \sim T^6,$ $p=\rho /5,$ called by Shuryak the 
``realistic'' equation of state for hot hadronic matter \cite{Shu}, which has
some experimental support. Zhirov and Shuryak \cite{ZS} have calculated 
the velocity of sound, $c_s^2\equiv dp/d\rho =c_s^2(T),$ with $p$ and $\rho $ 
defined in Eqs. (2),(3), and found that $c_s^2(T)$ at first increases with $T$
very quickly and then saturates at the value of $c_s^2\simeq 1/3$ if only the
pions are taken into account, and at $c_s^2\simeq 1/5$ if resonances up to 
$M\sim 1.7$ GeV are included. 

We have checked the coincidence of the results given by a linear spectrum (9) 
with those obtained directly from Eq. (2) for the actual hadronic species with
the corresponding degeneracies, for all well-established hadronic multiplets, 
\\ the mesons: 

1 $^3S_1$ $J^{PC}=1^{--}$ nonet, $\;\rho (770)\;,$ $\;\omega (783)\;,$ $\;\phi 
(1020)\;,$ $\;K^\ast (892)\;$ 

1 $^3D_3$ $J^{PC}=3^{--}$ nonet, $\rho _3(1690),$ $\omega 
_3(1670),$ $\phi _3(1850),$ $K^\ast _3(1780),$ \\ the baryons: 

$J^P=\frac{1}{2}^{+}$ octet, $\;N(939),$ $\Lambda (1116),$ $\Sigma (
1190),$ $\Xi (1320)$ 

$J^P=\frac{3}{2}^{+}$ decuplet, $\Delta (1232),$ $\Sigma (
1385),$ $\Xi (1530),$ $\Omega (1672)$ 

$J^P=\frac{3}{2}^{-}$ nonet, $N(1520),$ 
$\Lambda (1690),$ $\Sigma (1670),$ $\Xi (1820),$ $\Lambda (1520)$ 

$J^P=\frac{5}{2}^{+}$ octet, $N(1680),$ $\Lambda (1820),$ $\Sigma (1915),$ 
$\Xi (2030),$ \\ and found it excellent \cite{spectrum}. Therefore, the fact
established theoretically that a linear spectrum is the actual spectrum in the
description of individual hadronic multiplets, finds its experimental 
confirmation as well.

Now we wish to apply the linear spectrum (9) to the problem of the 
identification of the $q\bar{q}$ nonets, as follows: we shall be looking for 
such a composition of a nonet for which the results given by both the formulas
(2), and (4) with a linear spectrum, coincide (or, at least, are very close). 
We shall proceed as in a previous paper \cite{spectrum}: instead of a direct 
comparison of Eqs. (2) and (4), we shall compare the expressions $p/p_{SB}$ for
both cases, where $p_{SB}\equiv \sum _ig_i T^4/\pi ^2$ (i.e., $p_{SB}$ is the 
pressure in an ultrarelativistic gas with $g=\sum _ig_i$ degrees of freedom). 

In order to illustrate how the suggested method works in practice, let us first
consider another meson nonet not discussed above:

2 $^3S_1$ $J^{PC}=1^{--},$ $\;\rho (1450),$ $\omega (1390),$ $\phi (1680),$
$K^\ast (1410).$ \\
As suggested by the recent Particle Data Group \cite{data2}, in this currently
adopted $q\bar{q}$ assignment, the $K^\ast (1410)$ could be replaced by the
$K^\ast (1680)$ as the 2 $^3S_1$ state (The $K^\ast (1680)$ is currently placed
as a member of the 1 $^3D_1$ $J^{PC}=1^{--}$ nonet). We have checked the both 
possibilities, \\ 1) $\rho (1450),$ $\omega (1390),$ $\phi (1680),$
$K^\ast (1410),$ \\ 2) $\rho (1450),$ $\omega (1390),$ $\phi (1680),$
$K^\ast (1680),$ \\ and compared the ratios $p/p_{SB}$ in both cases. Our 
results are shown in Fig. 1,2. One sees that for the currently adopted 
assignment, the curves match very little; in contrast, for the assignment 
suggested by the Particle Data Group, the curves almost coincide. Therefore,
the latter is favored from the viewpoint of a hadronic resonance spectrum. 

Now we turn to the nonets discussed above. We have checked all possibilities 
for composing a nonet with the isoscalar states by choosing the latter among 
the states discussed above, and tried the three isovector states for the scalar
meson nonet. The following summarizes the results of this work. \\

1. For the scalar meson nonet, we found that with one of the states, $f_0(975)$
or $f_0(1710)$ included in the nonet, the curves $p/p_{SB}$ as calculated from
both Eqs. (2), and (4) with a linear spectrum, do not match, or match
very little, which means that the currently adopted nonet assignment (with the
$f_0(975)$ included) does not agree well with our criterion of a linear 
spectrum being the actual spectrum of a multiplet. With the remaining scalar 
mesons (among those discussed above) included, the following are
the combinations for which the curves coincide (Figs. 3--8):
\\ with the $a_0(980),$

1) $a_0(980),$ $f_0(1240),$ $f_0(1400),$ $K_0^\ast (1430),$ 

2) $a_0(980),$ $f_0(1400),$ $f_0(1525),$ $K_0^\ast (1430),$ 
     
3) $a_0(980),$ $f_0(1525),$ $f_0(1590),$ $K_0^\ast (1430);$ \\ 
with the $a_0(1320),$

4) $a_0(1320),$ $f_0(1240),$ $f_0(1525),$ $K_0^\ast (1430);$ \\ 
with the $a_0(1450),$

5) $a_0(1450),$ $f_0(1400),$ $f_0(1525),$ $K_0^\ast (1430).$ \\
The latter is the $q\bar{q}$ assignment suggested by Montanet \cite{Mont}. \\

2. For axial-vector and tensor meson nonets, in either case the best choice 
coincides with the composition currently adopted by the Particle Data Group, 
i.e.,

$a_1(1260),$ $f_1(1285),$ $f_1(1510),$ $K_{1A}$ \\ 
for the axial-vector nonet, and

$a_2(1320),$ $f_2(1270),$ $f_2^{'}(1525),$ $K_2^\ast (1430)$ \\ 
for the tensor nonet. \\ 

3. We have also checked the currently adopted composition of another nonet,

1 $^3F_4$ $J^{PC}=4^{++},$ $\;a_4(2040),$ $f_4(2050),$ $f_4(2220),$ $K_4^\ast
(2045),$ \\ of which two states, $a_4(2040)$ and $f_4(2220),$ need experimental
confirmation, and found that for this composition the curves match almost 
perfectly. \\ 

The main conclusion of this work is that a linear resonance spectrum suggests
a new $q\bar{q}$ assignment for the scalar meson nonet which favors the 
interpretation of the $f_0(975)$ and $f_0(1710)$ mesons as non-$q\bar{q}$ 
objects. As we have seen, a linear resonance spectrum turns out to work very 
effectively towards resolution of meson nonet enigmas, for all meson nonets 
discussed in the letter except for the scalar meson nonet for which it is 
still impossible to make a unique prediction. For the scalar meson nonet, 
further exclusion of non-$q\bar{q}$ states should be made by using the standard
methods, e.g., the Regge phenomenology. In this way, one may exclude the $a_0
(980)$ as a state which does not fit as a $q\bar{q}$ state lying on a linear 
Regge trajectory (as well as $f_0(975)$ and $f_1(1420)$ \cite{IMZ}), thus 
leaving two possibilities for the nonet,

$a_0(1320),$ $f_0(1240),$ $f_0(1525),$ $K_0^\ast (1430),$

$a_0(1450),$ $f_0(1400),$ $f_0(1525),$ $K_0^\ast (1430),$ \\
which may, in turn, be justified by the calculation of Horn and Schreiber 
\cite{HS} who calculated the mass of the lowest-lying $0^{++}$ scalar
state by using the $t$-expansion method for Hamiltonian lattice QCD with two 
massless quarks, to be $\sim 1.3$ GeV, which is close to the mass of 
the lowest-lying scalar state for these assignments.

One can speculate even further. The leading Regge trajectory corresponding to 
the $\rho $ meson resonances is described by the straight line
\beq
J=0.58+0.84\;M^2
\eeq
to a high accuracy. Six resonance states, the $\rho (770),$ $a_2(1320),$
$\rho _3(1690),$ $a_4(2040),$ $\rho _5(2350),$ $a_6(2450),$ with isospin
$I=1$ and $J^{PC}=1^{--},2^{++},3^{--},4^{++},5^{--},6^{++},$ respectively,
belong to this trajectory as the  1 $^3S_1,$ 1 $^3P_2,$ 1 $^3D_3,$ 1 $^3F_4,\;
\ldots $ states. In terms of the nonrelativistic quark model, these resonance
states are interpreted as the states of the spin $q\bar{q}$ triplet with the 
maximal angular momentum $L=0,1,2,3,4,5,$ respectively. The isovector states 
1 $^3P_0,$ 1 $^3D_1,$ 1 $^3F_2,\;\ldots $ should belong to the parallel 
daughter Regge trajectory and correspond to the states with $J^{PC}=0^{++},1^{
--},2^{++},\ldots ,$ and $L=0,1,2,\ldots ,$ respectively. Of these, the 
$\rho (1700)$ is well established as the 1 $^3D_1$ $J^{PC}=1^{--}$ state. 
Among the remaining two 1 $^3P_0$ $J^{PC}=0^{++}$ candidates, $a_0(1320)$ and 
$a_0(1450)$ [and $a_0(980)],$ the only $a_0(1320)$ belongs to the trajectory 
on which the $\rho (1700)$ lies, which is parallel to the leading one (it can 
be easily seen by observing that the $a_2(1320)$ and $\rho _3(1690),$ which 
belong to the latter, have the masses very close to those of the $a_0(1320)$ 
and $\rho (1700),$ respectively).

Analogously, the leading trajectory for the $\omega $ resonances is described 
by the straight line which almost coincide with (10),   
\beq
J=0.59+0.84\;M^2,
\eeq
and includes the resonance states 1 $^3S_1,$ 1 $^3P_2,$ 1 $^3D_3,\;\ldots $ 
with $I=0,$ the $\omega (783),$ $f_2(1270),$ $\omega _3(1670),$ $f_4(2050),$ 
$f_6(2510),$ having $J^{PC}=1^{--},2^{++},3^{--},4^{++},6^{++}$ and $L=0,1,2,
3,5$ respectively. The isoscalar states 1 $^3P_0,$ 1 $^3D_1,$ 1 $^3F_2,\;
\ldots ,$ should belong to the daughter trajectory as those with $J^{PC}=0^{
++},1^{--},2^{++},\ldots ,$ and $L=0,1,2,\ldots \;.$ Of these, the $\omega 
(1600)$ is well established as the 1 $^3D_1$ $J^{PC}=1^{--}$ state, and among 
the 1 $^3P_0$ $J^{PC}=0^{++}$ candidates, the only $f_0(1240)$ belongs to the 
same trajectory which is parallel to the leading one (it is easily seen by 
observing that the $f_2(1270)$ and $\omega _3(1670)$ belong to the latter).

Thus, the Regge phenomenology favors the only 1 $^3P_0$ $J^{PC}=0^{++}$ 
isovector $a_0(1320)$ and isoscalar $f_0(1240)$ states, leaving, therefore, 
the unique possibility for the scalar meson nonet assignment, from those 
already favored by a linear resonance spectrum,

$a_0(1320),$ $f_0(1240),$ $f_0(1525),$ $K_0^\ast (1430).$ \\
While the $K_0^\ast (1430)$ apparently belongs to the daughter trajectory,
$K_0^\ast (1430),$ $K^\ast (1680),$ $K_2^\ast (1980),$ which is parallel to the
leading one, $K^\ast (892),$ $K_2^\ast (1430),$ $K_3^\ast (1780),$ $K_4^\ast
(2045),$ $K_5^\ast (2380),$ described by the straight line 
\beq
J=0.32+0.84\;M^2,
\eeq
nothing can be said about the $f_0(1525),$ since its nearby trajectory partner,
the 1 $^3D_1$ $J^{PC}=1^{--}$ $s\bar{s}$ state, is not established at present.
We, however, note that the 1 $^3F_2$ $J^{PC}=2^{++}$ $s\bar{s}$ state may well
be the $f_2(2300)$ [or one of the $f_2(2340),f_2(2150),f_2(2175)],$ as
discussed above, which lies on a mutual trajectory with the $f_0(1525),$ 
parallel to the leading one, $\phi (1020),$ $f_2^{'}(1525),$ $\phi _3
(1850),$ $f_4(2220),$ described by the straight line 
\beq
J=0.04+0.84\;M^2.
\eeq 

Although our conclusions need experimental support, it is most probable that 
the true $q\bar{q}$ assignment for the scalar meson nonet is that for which 
there is a mass degeneracy of the scalar and tensor meson nonet states with
equal isospin and quark content, 

1 $^3P_0$ $J^{PC}=0^{++},\;\;$ $a_0(1320),$ $f_0(1240),$ $f_0(1525),$ 
$K_0^\ast (1430),$

1 $^3P_2$ $J^{PC}=2^{++},\;\;$ $a_2(1320),$ $f_2(1270),$ $f_2^{'}(1525),$ 
$K_2^\ast (1430).$

\newpage
\centerline{FIGURE CAPTIONS}
\bigskip
\bigskip
\bigskip
\bigskip
\hfil\break
Fig. 1. Temperature dependence of the ratio $p/p_{SB}$ as calculated from:
I. Eq. (2), II. Eq. (4) with a linear spectrum, for the 2 $^3S_1$ $J^{PC}=
1^{--}$ nonet with the assignment $\rho (1450),$ $\omega (1390),$ 
$\phi (1680),$ $K^\ast (1410).$\hfil\break
\hfil\break
\hfil\break
\hfil\break
Fig. 2. The same as Fig. 1 but with the $K^\ast (1410)$ replaced by $K^\ast
(1680).$\hfil\break
\hfil\break
\hfil\break
\hfil\break
Fig. 3. The same as Fig. 1 but for the scalar meson nonet with the assignment 
$a_0(980),$ $f_0(1240),$ $f_0(1400),$ $K_0^\ast (1430).$\hfil\break
\hfil\break
\hfil\break
\hfil\break
Fig. 4. The same as Fig. 3 but with the $f_0(1240)$ replaced by $f_0
(1525).$\hfil\break
\hfil\break
\hfil\break
\hfil\break
Fig. 5. The same as Fig. 3 but with the $f_0(1240),$ $f_0(1400)$ replaced by 
$f_0(1525),$ $f_0(1590).$\hfil\break
\hfil\break
\hfil\break
\hfil\break
Fig. 6. The same as Fig. 1 but for the scalar meson nonet with the assignment 
$a_0(1320),$ $f_0(1240),$ $f_0(1525),$ $K_0^\ast (1430).$\hfil\break
\hfil\break
\hfil\break
\hfil\break
Fig. 7. The same as Fig. 1 but for the scalar meson nonet with the assignment 
$a_0(1450),$ $f_0(1400),$ $f_0(1525),$ $K_0^\ast (1430).$\hfil\break
\hfil\break
\hfil\break
\hfil\break
Fig. 8. (An example of the curves which match little.) The same as Fig. 7
but with the $f_0(1525)$ replaced by $f_0(1590).$
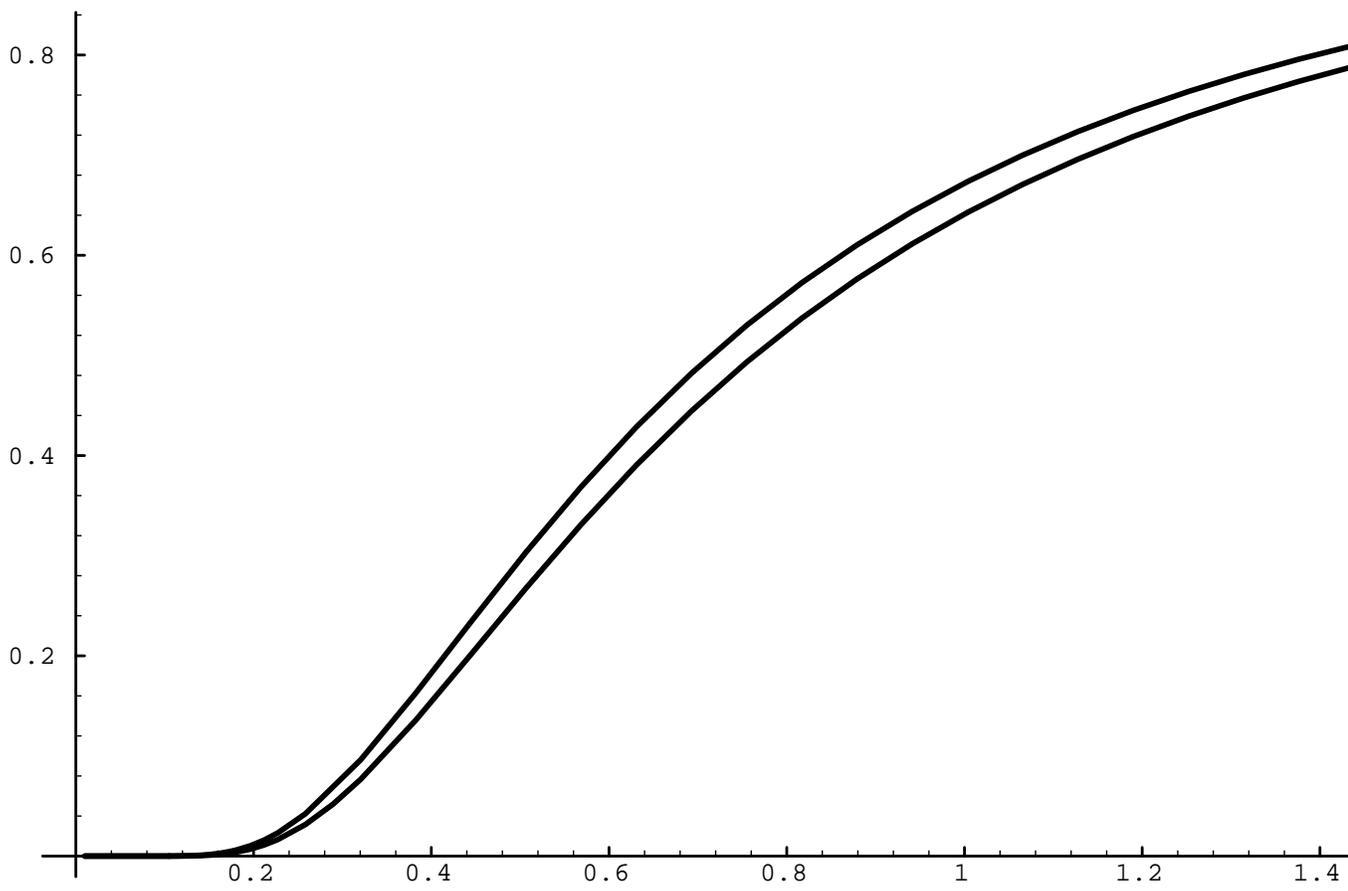
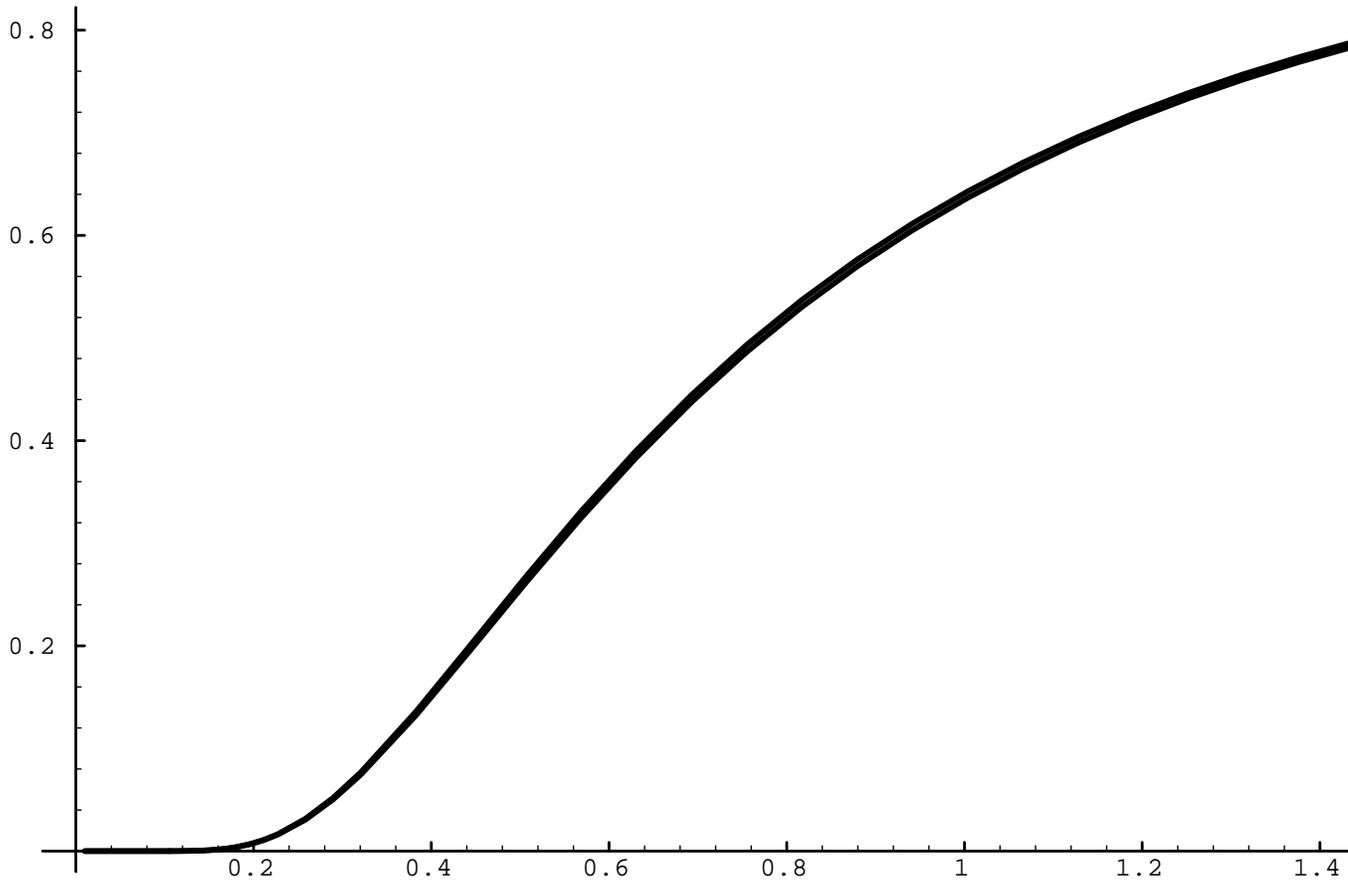
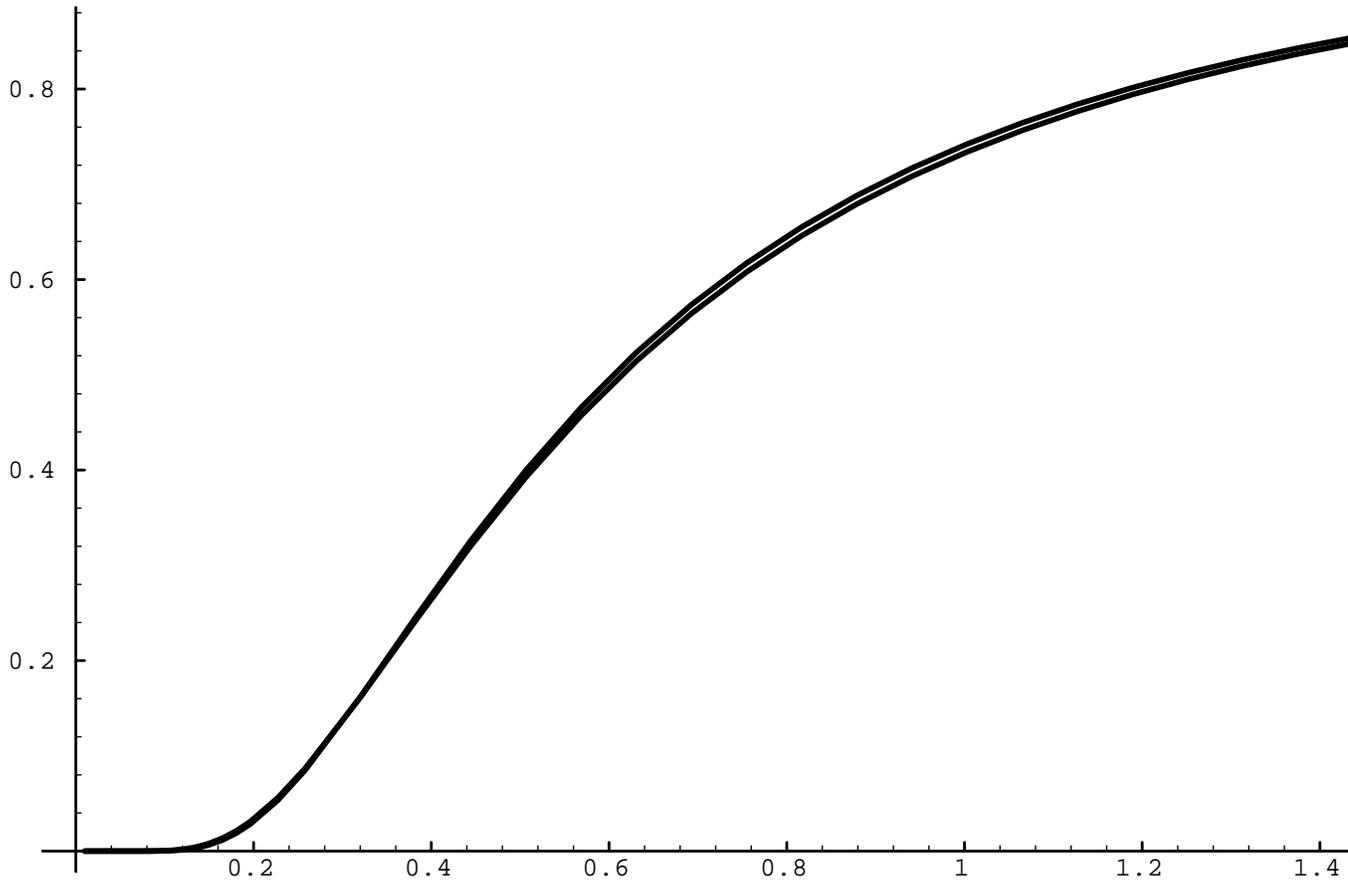
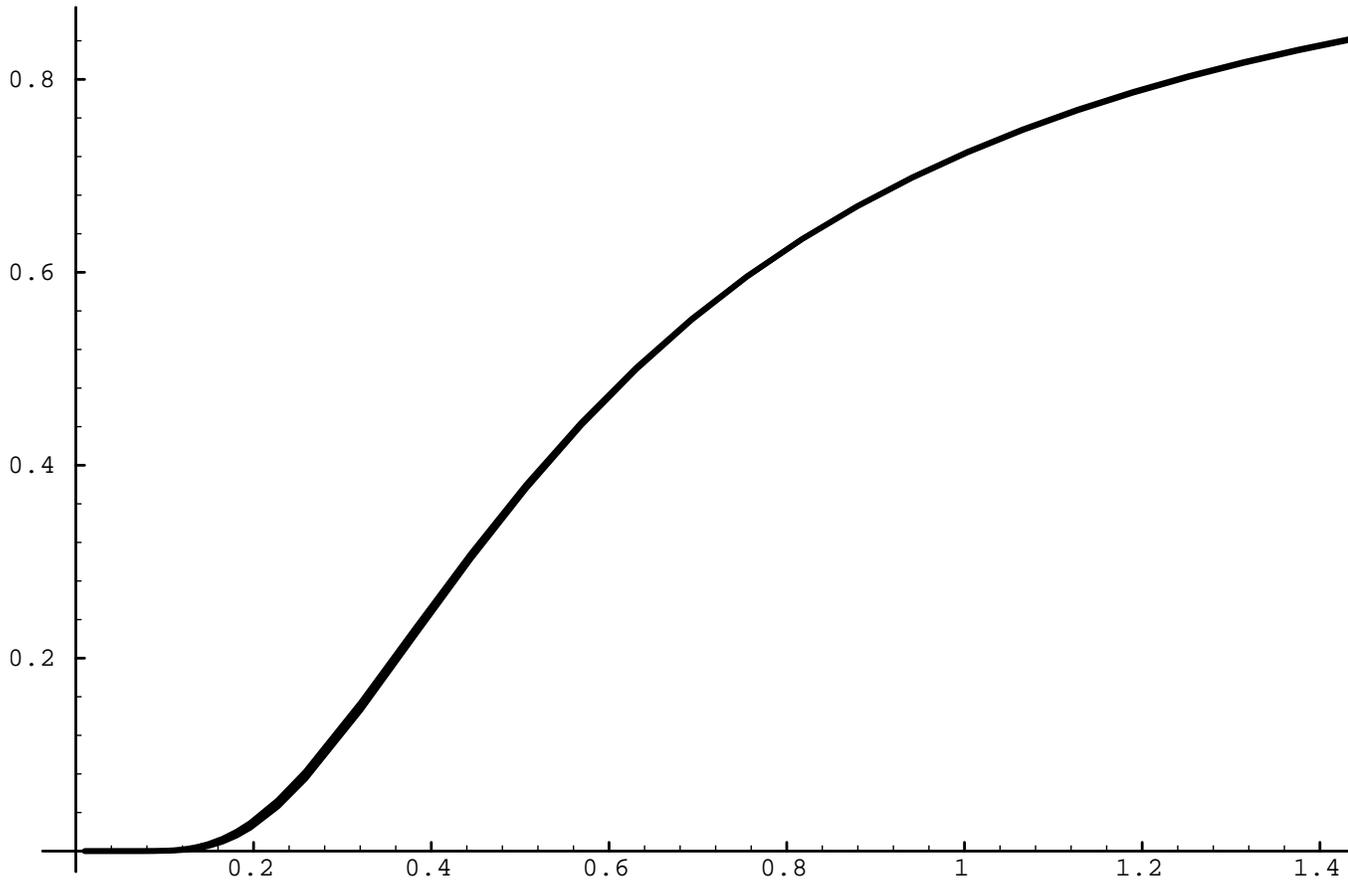
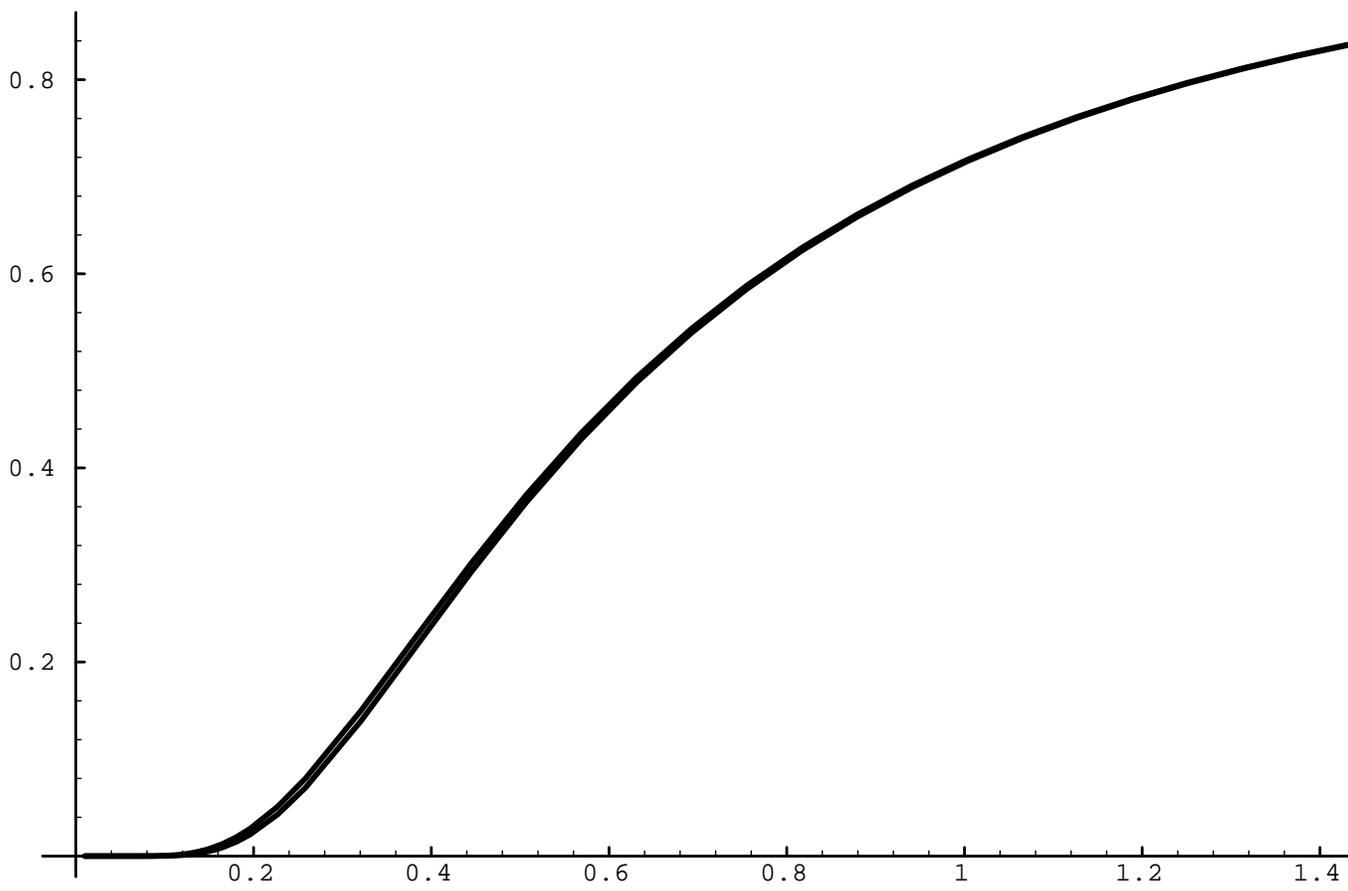
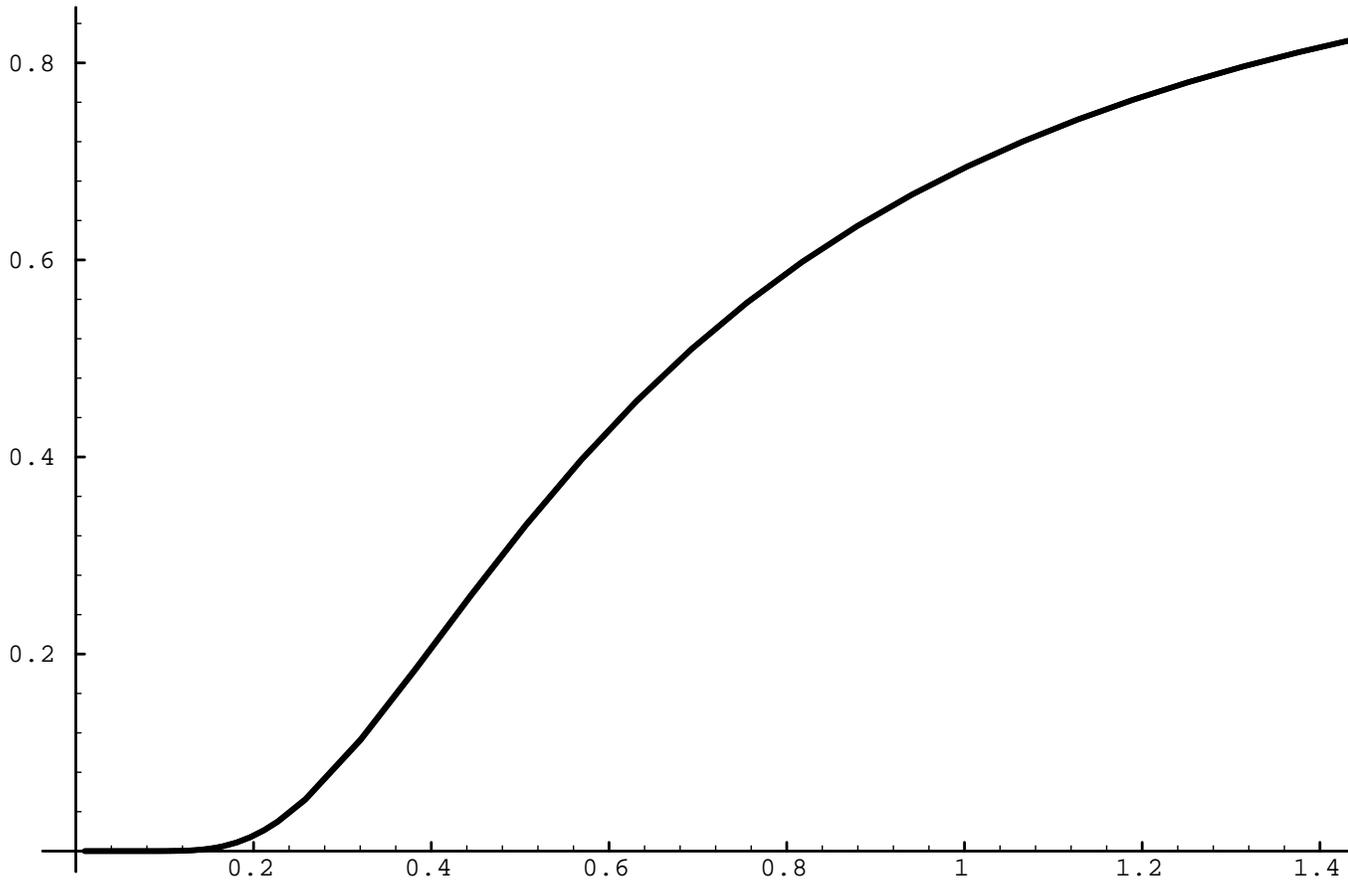
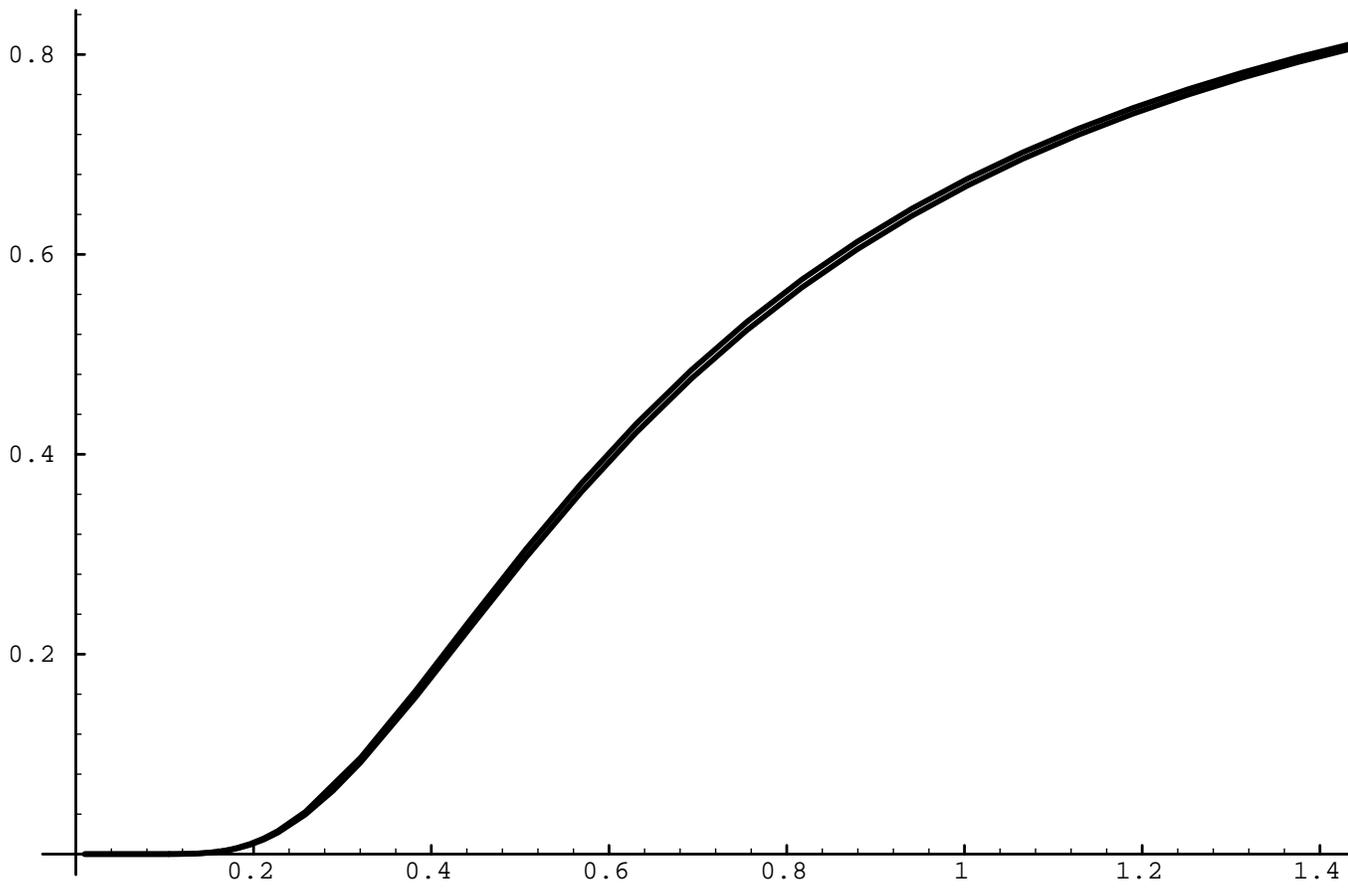
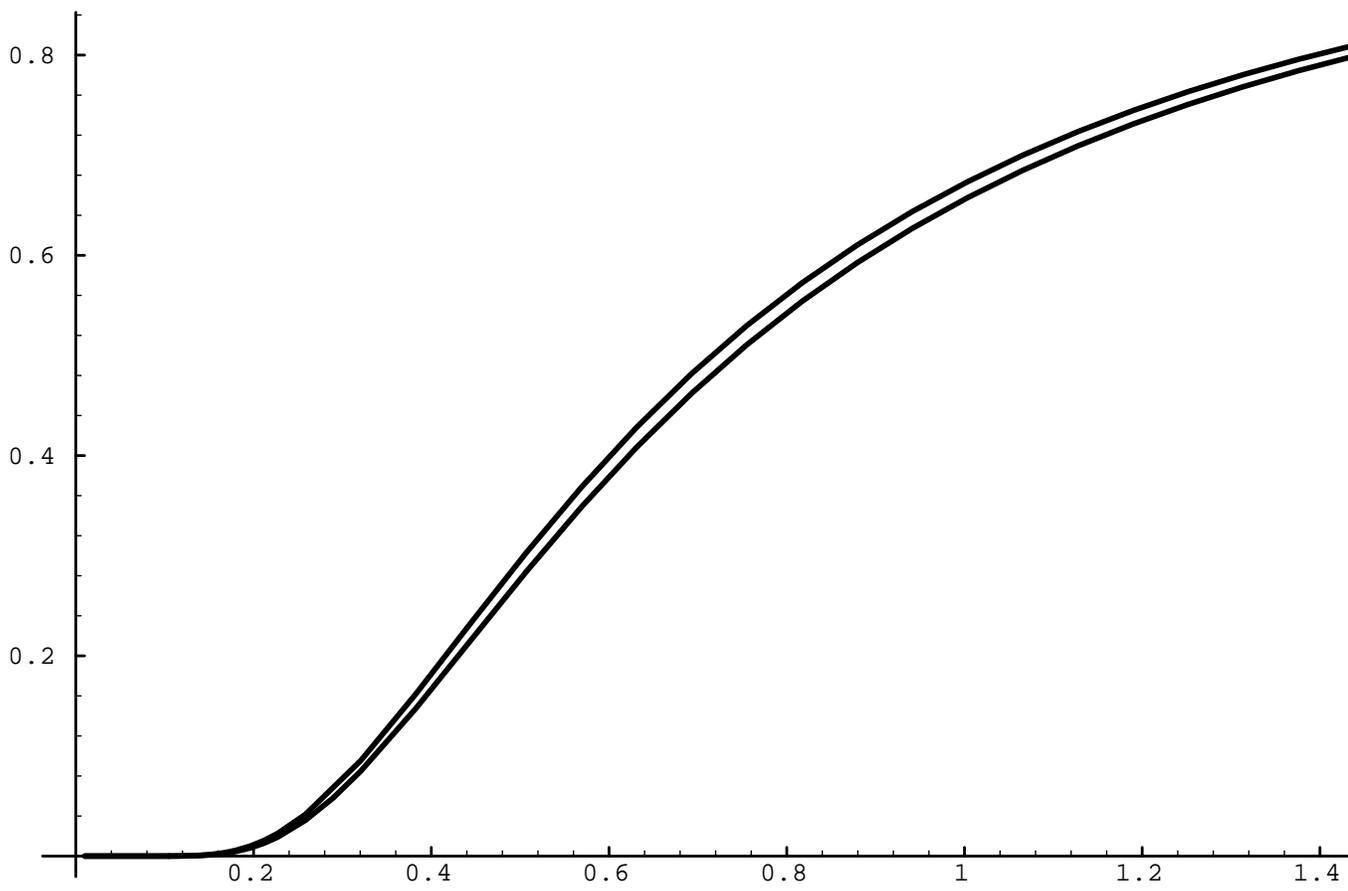
\end{document}